\documentclass[conference]{IEEEtran}
\usepackage{cite}
\usepackage{amsmath,amssymb,amsfonts}
\usepackage{algorithmic}
\usepackage{graphicx}
\usepackage{xcolor}
\usepackage{subfig}
\usepackage{makecell}
\usepackage{booktabs}
\usepackage{subfig}
\usepackage{tabularx}
\usepackage{float}

\usepackage{pgfplots}
\usepgfplotslibrary{groupplots,dateplot}
\usetikzlibrary{external,patterns,shapes.arrows}
\pgfplotsset{compat=newest}

\def\BibTeX{{\rm B\kern-.05em{\sc i\kern-.025em b}\kern-.08em
    T\kern-.1667em\lower.7ex\hbox{E}\kern-.125emX}}

\newcolumntype{Y}{>{\centering\arraybackslash}X}

\makeatletter
\def\footnoterule{\kern-3\p@
  \hrule \@width 2in \kern 2.6\p@} 
\makeatother

 \usepackage{acronym}

\acrodef{UAV}{Unmanned Aerial Vehicle}
\acrodef{EH}{Energy Harvesting}
\acrodef{MAF}{Moving Average Filter}
\acrodef{GPU}{Graphic Processing unit}
\acrodef{CPU}{Central Processing unit}

\newcommand{\review}[1]{\textcolor{blue}{#1}}
\renewcommand{\review}[1]{#1}

\begin{document}

\title{Accurate Calibration of Power Measurements from Internal Power Sensors on NVIDIA Jetson Devices}


\author{Neda Shalavi, Aria Khoshsirat, Marco Stellini, Andrea Zanella, Michele Rossi\\
Department of Information Engineering, University of Padova, Italy\\
Email: \texttt{\{name.surname\}@unipd.it}}

\IEEEoverridecommandlockouts

\maketitle
\makeatletter
\def\ps@IEEEtitlepagestyle{%
  \def\@oddfoot{\mycopyrightnotice}%
  \def\@oddhead{\hbox{}\@IEEEheaderstyle\leftmark\hfil\thepage}\relax
  \def\@evenhead{\@IEEEheaderstyle\thepage\hfil\leftmark\hbox{}}\relax
  \def\@evenfoot{}%
}
\def\mycopyrightnotice{%
  \begin{minipage}{\textwidth}
  \centering \scriptsize
  \copyright~2023 IEEE. Personal use of this material is permitted. Permission from IEEE must be obtained for all other uses, in any current or future media, including reprinting/republishing this material for advertising or promotional purposes, creating new collective works, for resale or redistribution to servers or lists, or reuse of any copyrighted component of this work in other works.
  \end{minipage}
}
\makeatother

\maketitle

\begin{abstract}
Power efficiency is a crucial consideration for embedded systems design, particularly in the field of edge computing and IoT devices. This study aims to calibrate the power measurements obtained from the \mbox{built-in} sensors of NVIDIA Jetson devices, facilitating the collection of reliable and precise power consumption data in real-time. To achieve this goal, accurate power readings are obtained using external hardware, and a regression model is proposed to map the sensor measurements to the true power values.
Our results provide insights into the accuracy and reliability of the \mbox{built-in} power sensors for various Jetson edge boards and highlight the importance of calibrating their internal power readings. In detail, internal sensors underestimate the actual power by up to $50\%$ in most cases, but this calibration reduces the error to within
$\pm 3\%$. By making the internal sensor data usable for precise online assessment of power and energy figures,
the regression models presented in this paper have practical applications, for both practitioners and researchers, in accurately designing energy-efficient and autonomous edge services.
\end{abstract}
\begin{IEEEkeywords}
Power measurements, IoT, Embedded systems, Edge Computing, NVIDIA Jetson.
\end{IEEEkeywords}

\section{Introduction}

Accurate energy analysis and power requirements are crucial aspects of embedded systems design, particularly in the field of edge computing and the Internet of Things (IoT). As the demand for autonomous and low-power computing devices grows, the need for accurate power measurement and management means becomes increasingly important. Results from~\cite{Jtsn_server} demonstrate that NVIDIA Jetson devices, specifically Jetson Xavier, provide a promising solution to edge-servers deployment or embedded edge applications due to their high peak performance per watt, especially in comparison with other edge platforms~\cite{comparison_jtsn_rspi}.
Energy consumption and efficiency have always been crucial aspects for \mbox{energy-constrained} and \mbox{battery-powered} devices in network edge scenarios~\cite{EE_survey}. The authors of~\cite{UAV_Jetson} proposed an \mbox{energy-aware} autonomous tracking and landing system for a \ac{UAV} equipped with an NVIDIA Jetson nano, exploiting the Jetson onboard power sensors to improve their solution. However, there have been debates~\cite{msrmnt_method} over the measurement methods and the precision of NVIDIA Jetson \mbox{built-in} sensors. Authors of~\cite{msrmnt_method} studied and proposed a valid method to calculate the energy consumption of \ac{GPU} kernel activities in NVIDIA Jetson Nano and AGX Xavier by only relying on \mbox{built-in} sensors. Similarly, authors of~\cite{Energy_profiler} conducted the power analysis and proposed a benchmark to profile the energy consumption of \mbox{energy-constrained} embedded applications on Jetson devices, but implementing their measurement setup is not practical for all researchers. In~\cite{obsrvd_gap_nano}, the Jetson Nano \mbox{built-in} input sensor for power measurement has been compared with a \mbox{real-time} power measurement from an external benchmark, observing a \mbox{non-negligible} gap between these two readings.

Authors of~\cite{obsrvd_gap_nano} argued that the reason for the observed gap between external and internal power measurements of Jetson Nano could be due to the power consumed by Jetson board elements between the input power plug and the whole \mbox{built-in} sensor on the board. However, \mbox{built-in} sensors typically measure average values, while failing to track current peaks. Furthermore, the sampling rate impacts the precision of the measured data~\cite{GPU_srvy}, and concurrently running programs impacts the \ac{CPU} access frequency to read from the onboard sensors, which, in turn, impacts the accuracy of the measured power. From our measurements, we do not think that the reason for this gap is attributable to the sampling frequency, as it is observed even when the power is nearly constant. Instead, it may be due to inefficiencies and losses within the power supply circuitry, or to systematic errors introduced by the built-in sensor measurement circuit.

Calibrating the power measurements of these sensors is key for ensuring the reliability of power measurement data and making informed decisions about power management. This is particularly important for tasks that require online power assessment, such as performance optimization and scheduling, energy-efficient computing, and battery-powered applications. Precise measurements are also required for the correct dimensioning of \ac{EH} systems for such scenarios where computing devices are combined with batteries and solar panels for self-sufficiency.
Authors of~\cite{ASU} proposed a framework to distribute computation between IoT devices (Jetson Nano) and a server aiming at minimizing the energy consumption of IoT devices by relying on power data measured from \mbox{build-in} sensors on the Jetson device. Authors of~\cite{tradeoff} demonstrated an energy-process-latency \mbox{trade-off} by scaling \ac{CPU}/\ac{GPU} frequency again relying on power data measured by NVIDIA Jetson TX2 sensors and plan to extend their work to optimize network resources. Another study~\cite{Aria} proposed splitting parallelizable tasks like video analytic/object detection among containers to minimize energy consumption and computation latency. They tested their proposal on Jetson TX2 and AGX Orin but only relied on \mbox{built-in} sensors for the power measurement data. Furthermore, authors of~\cite{Ahmed} studied the energy consumption of inference tasks on NVIDIA Jetson TX2 and Xavier as embedded edge devices using their \mbox{built-in} sensors and proposed an energy consumption estimation model based on the inference task parameters.

As we reviewed above, recently there has been an increasing interest in the literature to study or profile the energy consumption of various tasks on embedded IoT devices that require a \mbox{low-power} and autonomous operation, especially NVIDIA Jetson devices. Unfortunately, all these valuable studies that aim to provide energy consumption analysis at the edge with the aim of minimizing the energy consumption and edge network climate impacts or network resource management, did not have the opportunity to use actual power measurements of NVIDIA Jetson boards for more accurate energy analysis. 
Even though authors of~\cite{Energy_profiler, obsrvd_gap_nano} compared the \mbox{built-in} sensors measurement with power data measured by an external setup for specific cases, there is a lack of a general model or tool to acquire the real power data for NVIDIA Jetson devices. An accurate estimation of the power used by these devices at runtime is key to optimizing their performance and energy efficiency and ensuring their reliable operation in a variety of applications. We stress that, despite the growing importance of power measurements for such embedded systems, there has been limited research in this area, particularly with regard to characterizing and calibrating the power sensors of NVIDIA Jetson devices.
In this paper, we use an external power measurement setup to derive calibration functions for the power readings acquired from \mbox{built-in} power sensors of four commercially available NVIDIA Jetson devices, namely, Jetson AGX Orin, Jetson Xavier NX, Jetson TX2, and Jetson Nano. We characterize the gap between power measures from the internal sensor and accurate data acquired by an external oscilloscope and derive a simple analytical expression to correct it, achieving maximum errors of $\pm 3\%$ after our correction is applied. \review{The presented model can be useful in real-world edge network scenarios, allowing the online estimation of energy consumption figures for the implementation of resource management solutions~\cite{HiTDL}.}

\section{Methodology}

\begin{figure}[t]
\centerline{\includegraphics[width=0.4\textwidth]{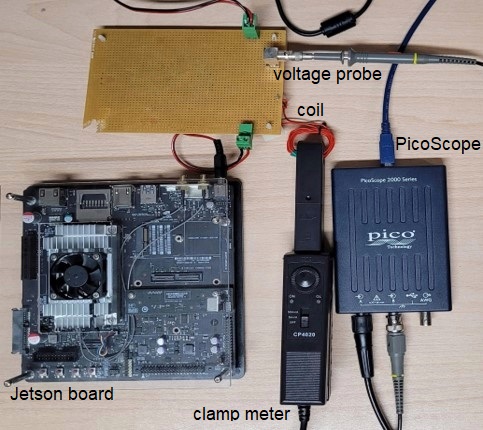}}
\centering
\caption{The external measurement setup.}\vspace{-0.3cm}
\label{fig:setup_image}
\end{figure}

The objective of this study is to map the \mbox{built-in} sensors' power measurements of NVIDIA Jetson devices to external accurate power measurements via a simple analytical model. To achieve this, the following methodology is employed:

\begin{itemize}

    \item \textbf{Hardware setup:} As indicated in~\cite{GPU_srvy}, the sampling rate of the power measurements directly impacts their precision for \mbox{\ac{GPU}-enabled} edge computers. The hardware used in this study consists of a set of commercially available NVIDIA Jetson devices and a high sampling rate external oscilloscope to accurately measure, store and stream power consumption data; the measurement setup is depicted in Fig.~\ref{fig:setup_image}.

    \item \textbf{Data collection:} We use a separate thread to read data at the maximum possible rate from \mbox{built-in} sensors without interruption, while another thread runs various workloads on the Jetson device to create a diverse range of power levels. The external power meter simultaneously records the total power measurement of the edge computer. As confirmed by~\cite{msrmnt_method}, \ac{MAF} is useful to reduce white noise on power measurements from computing devices, specifically Jetson devices. We applied \ac{MAF} on sampled data over $100$~ms for both measured data types (internal sensor and external oscilloscope) to reduce the measurement noise before any further processing. Power measurement data is collected by running various workloads on the Jetson devices, while simultaneously reading the power consumption from the built-in power sensor and from the external power meter.

    \item \textbf{Mathematical model:} Based on the collected data, a simple regression model is fine-tuned to describe the relationship between the readings from the built-in power sensors and those from the external power meters. This model can be utilized to calibrate the readings from the built-in power sensors, improving the power measurements' accuracy.

    \item \textbf{Validation:} The calibrated readings from the built-in power sensors are validated, by comparing them to those from the external power meter. The accuracy of the calibrated readings is evaluated using the mean absolute error metric, gauging the maximum error for each device. 

    \item \textbf{Results and discussion:} The results are analyzed and discussed in terms of the accuracy of the calibrated readings from the built-in power sensors, and the overall effectiveness of the proposed calibration method.
\end{itemize}

Through this methodology, we show that, although the power measurements from the internal sensors of NVIDIA Jetson devices return highly inaccurate data, their calibration leads to satisfactory results, achieving errors of $\pm 3\%$ in the worst case with respect to accurate readings from an external measurement setup.

\section{Experimental Setup}

\begin{figure}[t]
\centerline{\includegraphics[width=0.4\textwidth,height=2in]{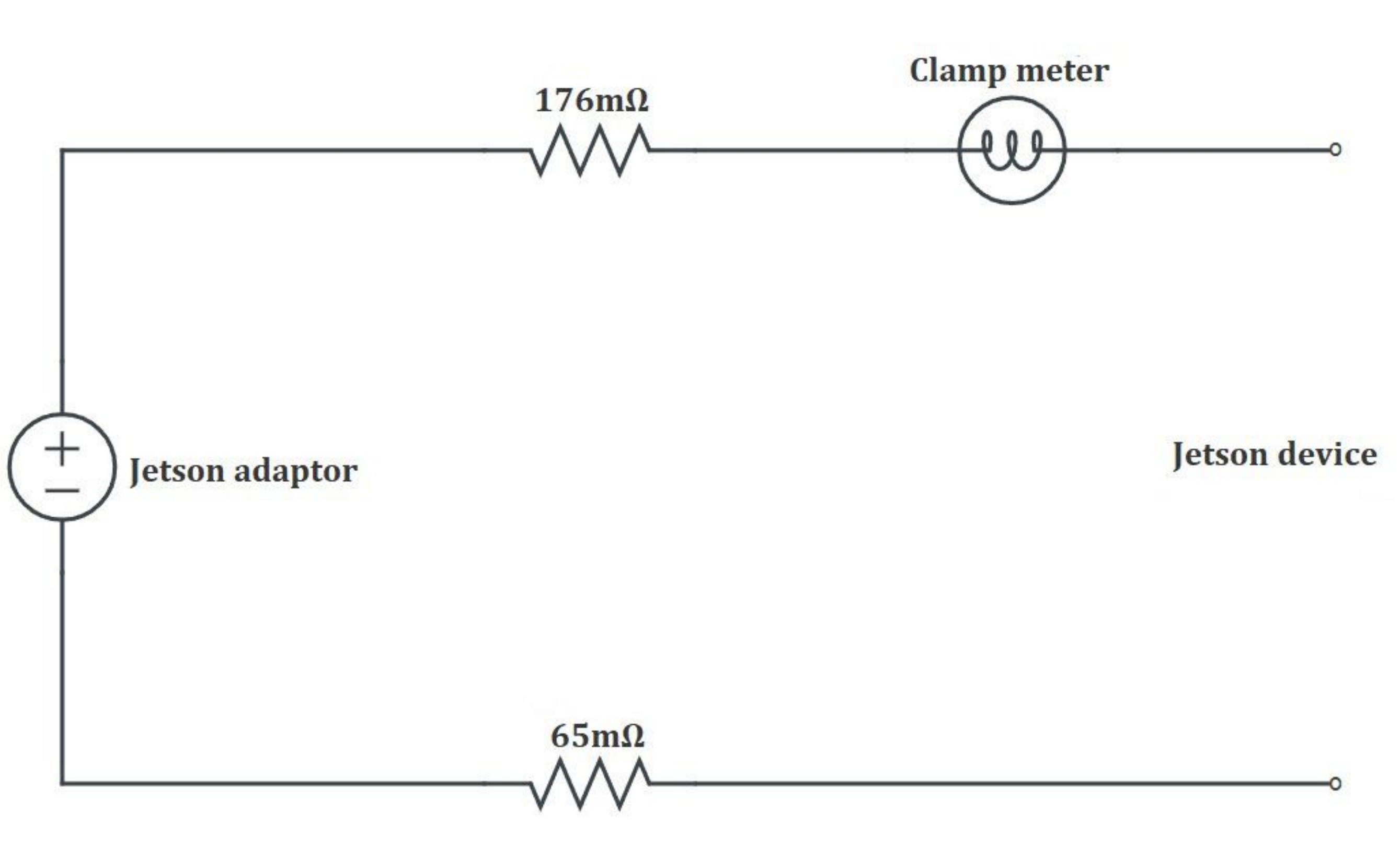}}
\centering
\caption{Measurement setup equivalent circuit.}\vspace{-0.3cm}
\label{fig:setup}
\end{figure}

Power measurements can be obtained from the Jetson devices via the onboard INA3221 sensors. By reading the \emph{sysfs} nodes, the power values of the whole board are extracted from the power sensor in our experiments. In the case of Jetson AGX Orin, where a whole board sensor is absent, the power sensor values of individual parts of the device are summed up to estimate the overall power value for the board.

For the external measurements, we utilized the PicoScope 2204A, a USB PC oscilloscope with $100$~MS/sec \review{(Mega Samples/sec)} sampling frequency, $10$~MHz of bandwidth, $8$ bit ADC resolution, two measurement channels, and a buffer memory of up to $128$~MS to stream measured data. We recorded data using the PicoScope streaming mode with a sampling interval set to $1$~ms. We recorded network time with \mbox{micro-second} precision alongside each recorded sample on PicoScope and \mbox{built-in} sensors for an accurate alignment of the data. The PicoScope has two channels that can measure input voltage and current simultaneously. To measure the input current and voltage of the Jetson devices from their adapters, we designed an inline measurement setup using \mbox{ad-hoc} connectors to place the measurement circuit in series between the adapter and the Jetson computer. Jetson adapters provide DC current and voltage to the device. Every DC adapter has red and black wires as positive (ungrounded) and negative (grounded) conductors. We used the Siglent CP4020 current probe (clamp meter) to sample the input current flowing into the Jetson device and added a \mbox{ten-round} coil in series (the red wire on which the clamp is attached in Fig.~\ref{fig:setup_image}) to increase the measurement precision. The added red wire has an impedance of $176$~m$\Omega$, and the added black wire has an impedance of $65$~m$\Omega$, therefore the impact of the measurement setup is negligible. Fig.~\ref{fig:setup} depicts the schematic diagram of the measurement circuit, where the inline measurement equivalent components are in series with the adapter and the Jetson devices, and the impedance values of the added red and black wires are shown. \review{Additionally, ensuring the accuracy and calibration of the external measurement setup is of utmost importance. In our work, we employed advanced equipment such as the \mbox{DSO-X 3024A} digital storage oscilloscope and the Keysight 1147B high precision current probe to verify the accuracy of our measurements.} 

\begin{figure}[t]
\centerline{\includegraphics[width=0.4\textwidth,height=2in]{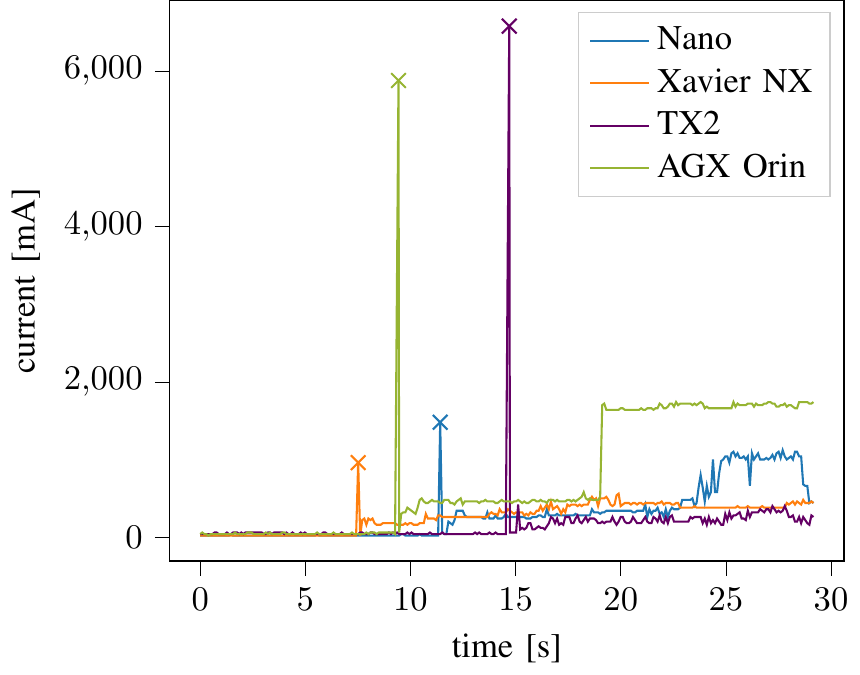}}
\centering
    \caption{Start up peak current of Jetson devices.}\vspace{-0.3cm}
    \label{fig:peak_current}
\end{figure}

For our experiments, we measured the power, using the sensors and hardware setup, while running different tasks on the Jetson devices. These tasks include the idle state of the operating system (with and without connecting a monitor), browsing the Internet from application windows, high-definition Video streaming, inference of various neural network models using \ac{GPU}, and running multiple instances of Python applications in parallel.

\section{Results}
\label{sec:results}


\begin{figure*}[ht!]
    \centering
    \subfloat[Nano]{\resizebox{.25\textwidth}{!}{\includegraphics[width=1\textwidth]{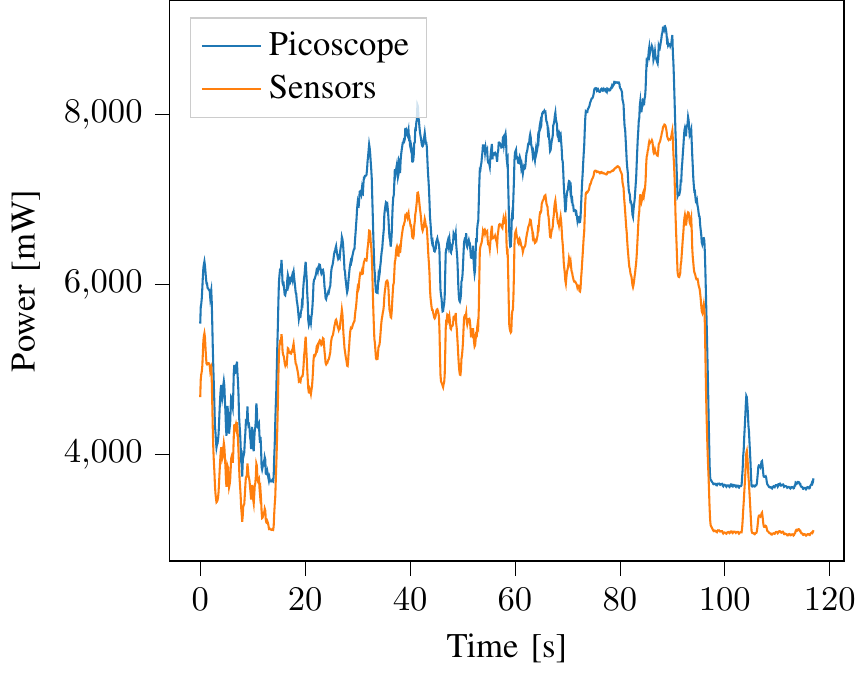}}}
    \subfloat[TX2]{\resizebox{.25\textwidth}{!}{\includegraphics[width=1\textwidth]{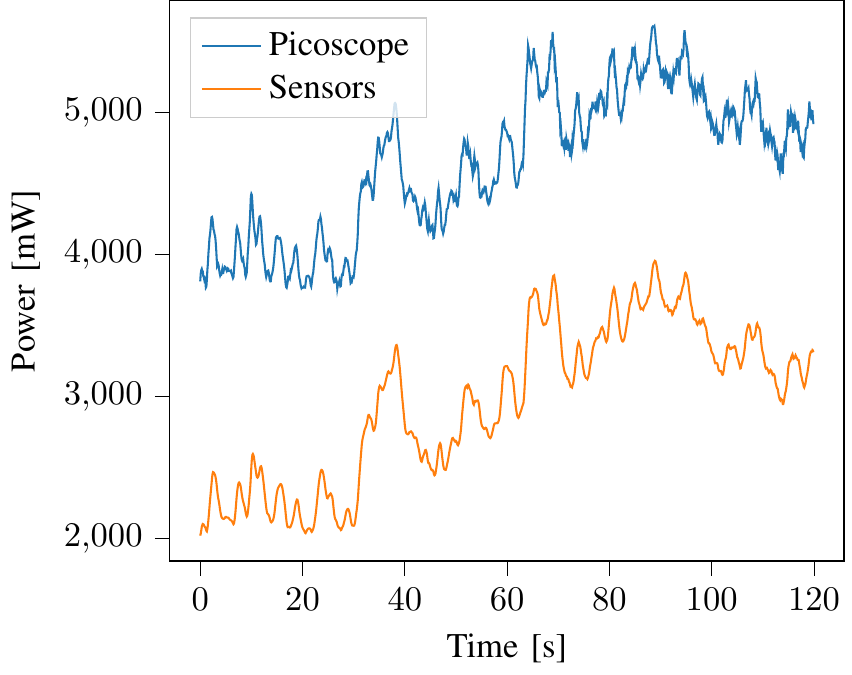}}}
    \subfloat[AGX Orin]{\resizebox{.24\textwidth}{!}{\includegraphics[width=1\textwidth]{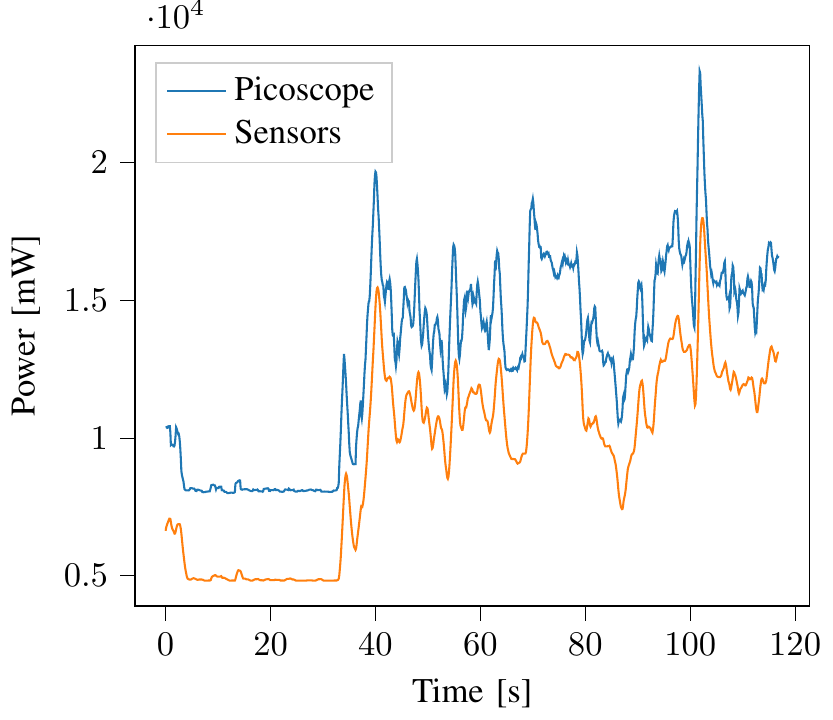}}}
    \subfloat[Xavier NX]{\resizebox{.24\textwidth}{!}{\includegraphics[width=1\textwidth]{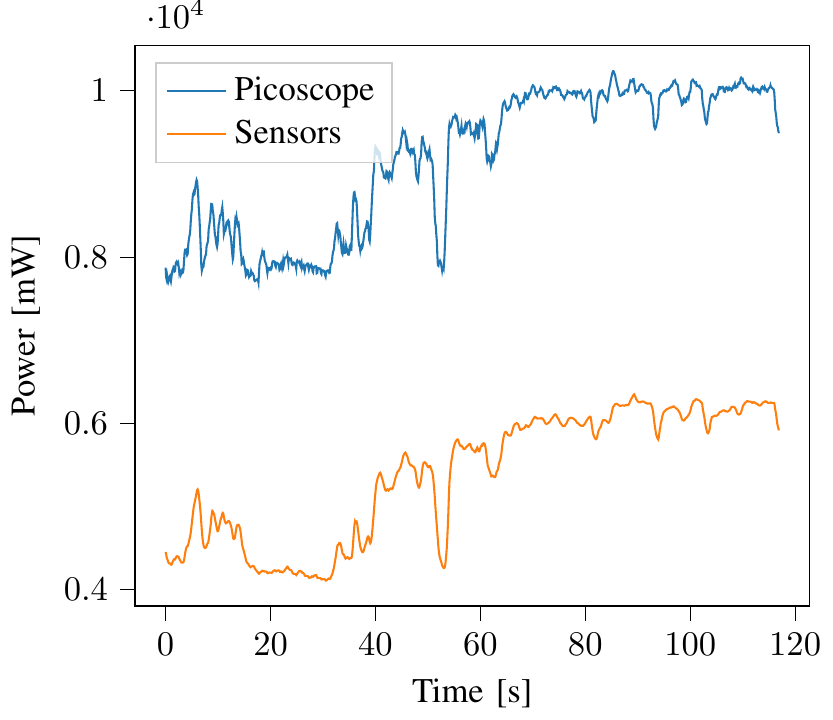}}}
    \caption{Power gap between external and \mbox{built-in} sensor measurements.}
    \label{samples}
\end{figure*}

\begin{figure*}[ht!]
     \centering
    \subfloat[Nano]{\resizebox{.25\textwidth}{!}{\includegraphics[width=1\textwidth]{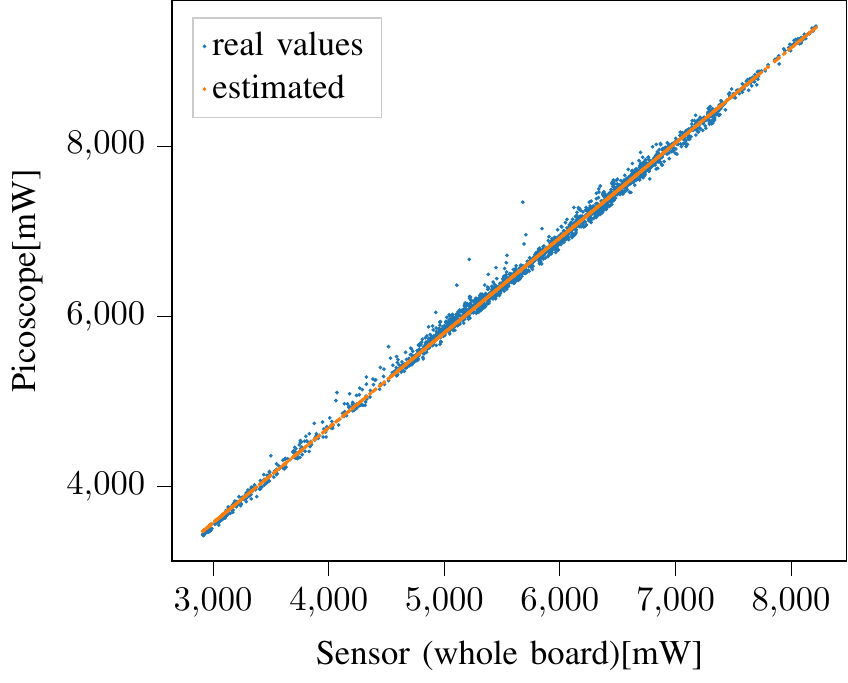}}}
    \subfloat[TX2]{\resizebox{.25\textwidth}{!}{\includegraphics[width=1\textwidth]{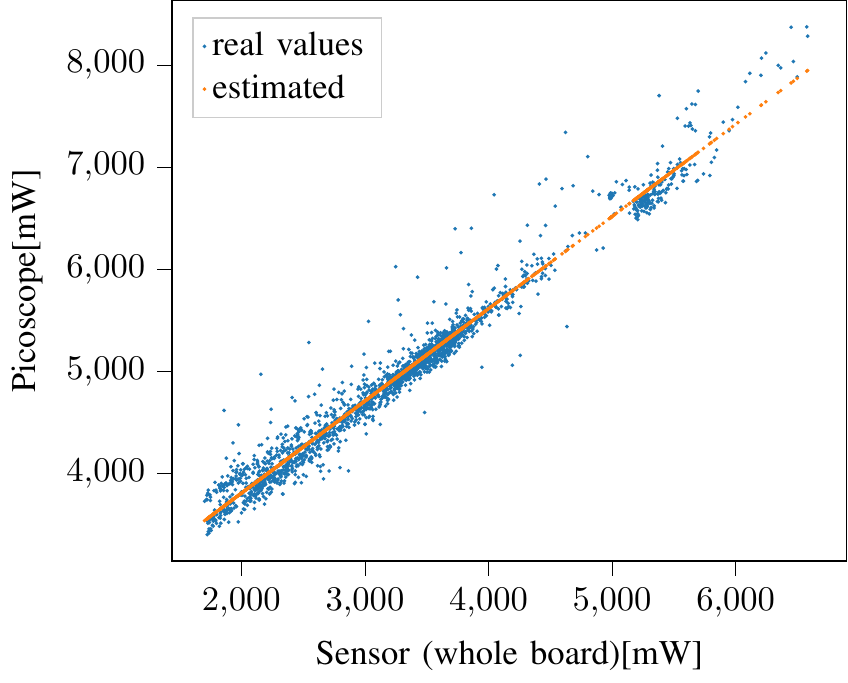}}}
    \subfloat[AGX Orin]{\resizebox{.24\textwidth}{!}{\includegraphics[width=1\textwidth]{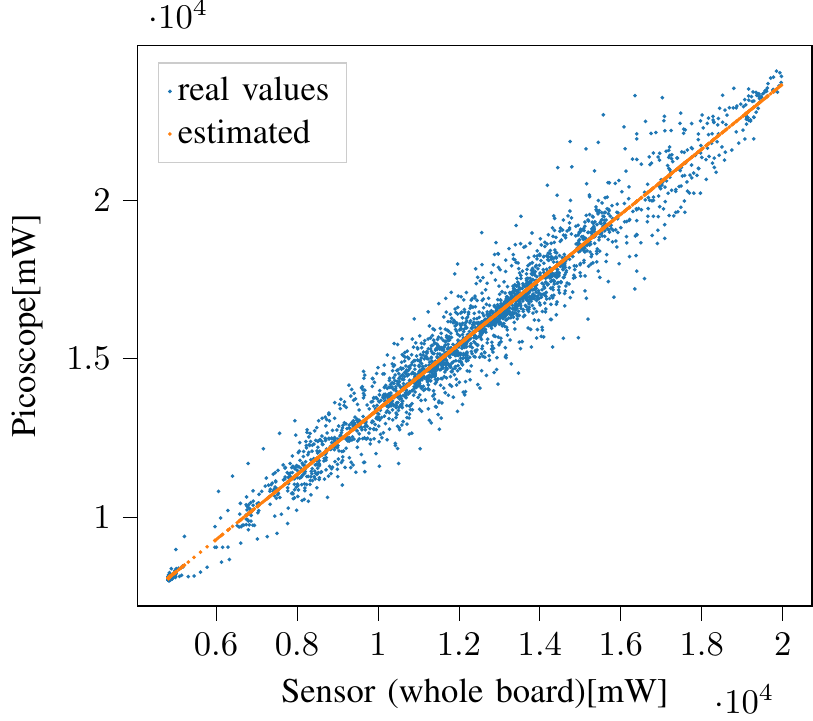}}}
    \subfloat[Xavier NX]{\resizebox{.24\textwidth}{!}{\includegraphics[width=1\textwidth]{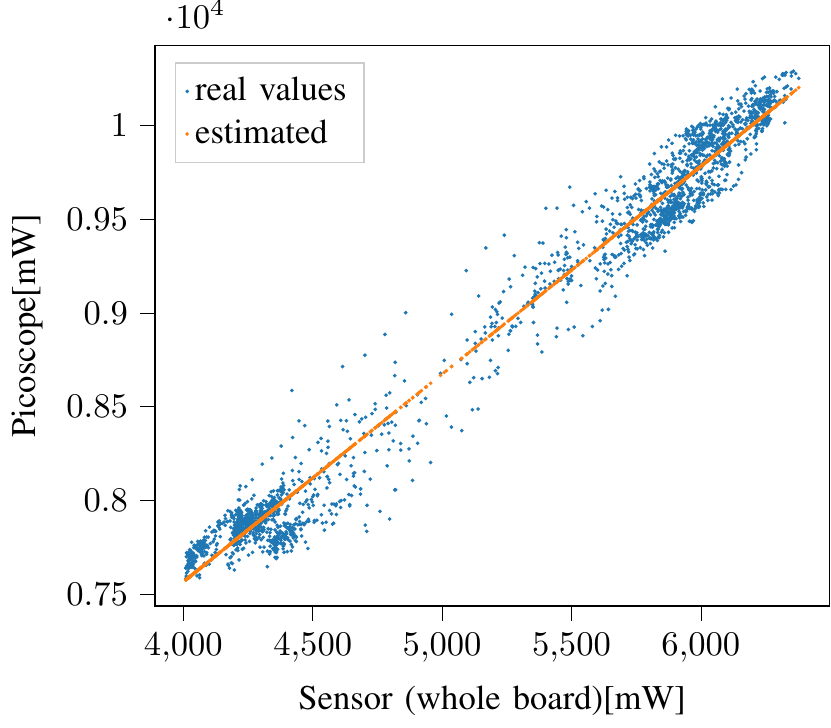}}}
     \caption{Linear regression mapping between internal (built-in sensor) and external (oscilloscope) power readings.}
     \label{fig:regression}
\end{figure*}

Attaining a realistic view of the power requirements of edge devices is necessary to choose the right battery size or \ac{EH} power supply. As a preliminary characterization, we have measured the power required to turn on the devices. For this, we observed that all Jetson boards require a high peak current in a very short time interval at boot time. 
Fig.~\ref{fig:peak_current} presents the current before and after turning on the devices. The current peak has a very short duration of less than $1$~ms -- if the battery can not provide such peak current, it fails to boot the Jetson device. The current peak for Jetson AGX Orin, TX2, Xavier NX, and Nano is $5880$~mA, $6580$~mA, $960$~mA, and $1480$~mA, respectively. Note that this peak current can only be measured via an external measurement setup, as the edge device is not yet in the position of providing power data during this transient phase.

Hence, we compared the power readings from the internal sensors with those gathered from the external oscilloscope, observing that the gap between the two differs for each Jetson device. In Fig.~\ref{samples}, we present a sample trace of our power measurements for each device, taken over a period of two minutes. Our experiments indicate that the gap between the internal power measurements and the external readings is {\it significant} (up to $50$\% of the actual power is missing from the internal board values), but the envelope of the power signal is preserved. Also, the gap remains nearly constant for the same device, regardless of the power values being measured.

Thus, we collected datasets from whole board sensors of Jetson devices and used regression to find suitable mapping functions. Fig.~\ref{fig:regression} presents the attained regression model, revealing a {\it linear} relationship between the \mbox{built-in} sensor measurements and the external measurements for Jetson Nano, TX2, AGX Orin, and Xavier NX (from left to right). This linear relationship can be used to map the power readings from the internal sensor into the actual power drained by the edge computer, as it would be measured by a calibrated and highly accurate external oscilloscope. The accuracy of such mapping was calculated by gauging the percentage error of such estimates with respect to the external power readings. The results show that the proposed mappings achieve accurate results for all tested devices. 

The mapping functions and their corresponding errors are given in the below Table~\ref{table:models}, reporting the formulas used to estimate the actual power from the built-in sensor readings (denoted by $x$ and expressed in milli-Watt), for each edge device. The percentage error represents the mean absolute error between the predicted power estimates and the external measurements from the oscilloscope. \review{Researchers can readily utilize these models to infer, at runtime, power figures of Jetson devices by using the readings obtained from their \mbox{built-in} sensors.}

\begin{table}[ht!]
\centering
\caption{Regression mapping models for Jetson devices}
\label{tab:model_parameters}
\begin{tabularx}{.48\textwidth}{lcYcY}
\toprule
       & \textbf{Device} & \textbf{Power calibration model [mW]} &  \textbf{Error} \\ \midrule\midrule
&\textbf{AGX Orin} &   $1.02 x + 3115.39$        & $ \pm 3 \%$ \\
&\textbf{Xavier NX} &      $1.10 x + 3130.41$     &             $\pm 2 \%$ \\
&\textbf{TX2}  &      $0.90x + 1998.80$     &              $\pm 3 \% $\\
&\textbf{Nano} &    $1.11 x + 232.60 $      &              $\pm 0.8 \%$\\ \bottomrule
\end{tabularx}
\label{table:models}
\end{table}

By mapping the built-in sensor measurements to the more accurate external measurements, we can perform a more reliable and accurate assessment of each device's performance. The proposed approach provides a simple and efficient solution for mapping built-in sensors to external sensors in autonomous edge systems.

\section{Concluding Remarks}

Accurate power measurements are key for optimizing performance and energy efficiency in embedded systems, particularly in the field of \ac{EH} edge computing and IoT devices. This study focused on calibrating the power measurements gathered from \mbox{built-in} NVIDIA Jetson device sensors, so as to obtain accurate power estimates. Our methodology involved using external and internal measurement hardware/software, and regression models to map the internal power measurements via linear functions. These functions can be utilized in practical \ac{EH} embedded setups with NVIDIA Jetson devices, or with simulated resource optimization algorithms, constituting a useful tool for practitioners and researchers who would like to carry out investigations on energy-ware resource allocation, and energy-efficient processing functions. Furthermore, our results provided insights into the accuracy and reliability of the built-in power sensors of various Jetson edge computers, highlighting the importance of calibrating power readings from the internal sensors. Our findings contribute to the field of power measurement in embedded systems and have implications for energy-efficient and autonomous device design. We aim to expand our work by attaining realistic energy consumption models of machine learning algorithms, making it possible to schedule computing resources at runtime in an energy-aware fashion. 

\section*{Acknowledgment}
This work has been supported by the EU H2020 MSCA ITN project Greenedge (grant no. 953775), and by the EU under the Italian ``Piano Nazionale di Ripresa e Resilienza'' (PNRR) of NextGenerationEU, partnership on ``Telecommunications of the Future'' (PE0000001 - program ``RESTART'').

\vspace{12pt}
\bibliographystyle{IEEEtran}
\bibliography{ms}
\end{document}